%% file: taulll_prl.tex
\documentclass[twocolumn,showpacs,aps,prl,superscriptaddress]{revtex4}

\usepackage{graphicx}
\usepackage{dcolumn}
\usepackage{amsmath}
\usepackage{epsfig}

\input pubboard/babarsym.tex

\newcommand{\lumi}    {376\invfb}
\newcommand{\lumion}  {339\invfb}
\newcommand{\lumioff} {37\invfb}

\newcommand{\lll}     {\ensuremath{\ell^{-}\ell^{+}\ell^{-}}}
\newcommand{\eee}     {\ensuremath{e^-\!e^+\!e^-}}
\newcommand{\eemw}    {\ensuremath{\mu^+\!e^-\!e^-}}
\newcommand{\eemr}    {\ensuremath{\mu^-\!e^+\!e^-}}
\newcommand{\emmw}    {\ensuremath{e^+\!\mu^-\!\mu^-}}
\newcommand{\emmr}    {\ensuremath{e^-\!\mu^+\!\mu^-}}
\newcommand{\mmm}     {\ensuremath{\mu^-\!\mu^+\!\mu^-}}

\newcommand{\taulll}  {\ensuremath{\tau^{-}\!\to\lll}}
\newcommand{\dEdM}    {\ensuremath{(\Delta E, \Delta M)}}
\newcommand{\dMdE}    {\ensuremath{(\Delta M, \Delta E)}}

\def\kk2f       {\mbox{\tt KK2f}\xspace}
\def\tauola     {\mbox{\tt Tauola}\xspace}
\def\photos     {\mbox{\tt Photos}\xspace}
\newcommand{\Nobs}      {\ensuremath{N_{\rm obs}}}
\newcommand{\Nbgd}      {\ensuremath{N_{\rm bgd}}}
\newcommand{\Nul}       {\ensuremath{N_{\rm UL}^{90}}}
\newcommand{\BRul}       {\ensuremath{\BR_{\rm UL}^{90}}}

\newcommand{\BABARPubYear}    {07}
\newcommand{\BABARPubNumber}  {059}
\newcommand{\SLACPubNumber}   {12766}

\def\figurebox#1#2#3{%
    \def\arg{#3}%
    \ifx\arg\empty
    {\hfill\vbox{\hsize#2\hrule\hbox to #2{\vrule\hfill\vbox to #1{\hsize#2\vfill}\vrule}\hrule}\hfill}%
    \else
    {\hfill\epsfbox{#3}\hfill}%
    \fi}

\begin{document}

\preprint{\babar-PUB-\BABARPubYear/\BABARPubNumber} 
\preprint{SLAC-PUB-\SLACPubNumber} 

\begin{flushleft}
\babar-PUB-\BABARPubYear/\BABARPubNumber\\
SLAC-PUB-\SLACPubNumber\\
\end{flushleft}

\title{
{\large \bf \boldmath
Improved Limits on the Lepton-Flavor Violating Decays \taulll}
}

\input pubboard/authors_jul2007.tex

\date{\today}

\begin{abstract}
A search for the neutrinoless, lepton-flavor violating decay of the 
tau lepton into three charged leptons has been performed 
using \lumi\ of data collected at an \epem\
center-of-mass energy around 10.58\gev with the \babar\ detector 
at the \pep2\ storage rings.
In all six decay modes considered, the numbers of events found 
in data are compatible with the background expectations.
Upper limits on the branching fractions are set in the range 
$(4-8) \times10^{-8}$ at 90\% confidence level.
\end{abstract}

\pacs{13.35.Dx, 14.60.Fg, 11.30.Hv}

\maketitle

Lepton-flavor violation (LFV) involving charged leptons has 
never been observed, and stringent experimental limits 
exist from muon branching fractions:
$\BR(\mmu\to\electron\gamma) < 1.2 \times 10^{-11}$ \cite{brooks99}
and $\BR(\mmu\to\electron\electron\electron) < 1.0 
\times 10^{-12}$ \cite{sindrum88} at 90\% confidence level (CL).
Recent results from neutrino oscillation experiments \cite{neut} 
show that LFV does indeed occur, although the branching fractions 
expected in charged lepton decays due to neutrino mixing alone 
are probably no more than $10^{-14}$ \cite{pham98}.

In tau decays, the most stringent limit on LFV is 
$\BR(\mtau\to\mu\gamma) < 4.5 \times 10^{-8}$ 
at 90\% CL \cite{belle07}.
Many descriptions of physics beyond the Standard Model (SM), 
particularly models 
seeking to describe neutrino mixing, predict enhanced LFV in tau 
decays over muon decays with branching fractions from 
$10^{-10}$ up to the current experimental limits 
\cite{paradisi05, babu02, brignole03}.
An observation of LFV in tau decays would be a 
clear signature of non-SM physics, while improved 
limits will provide further constraints on theoretical models.

This paper presents a search for LFV in the neutrinoless
decay \taulll{}, where $\ell$ is an electron or muon.
All possible lepton combinations consistent with charge
conservation are considered, leading to six distinct
decay modes (\eee, \eemw, \eemr, \emmw, \emmr, \mmm)
\cite{cc}.
The analysis is based on data recorded 
by the \babar\ detector at the \pep2\ asymmetric-energy \epem\ 
storage rings operated at the Stanford Linear Accelerator Center.
The data sample consists of \lumion\ recorded at
$\sqrt{s} = 10.58 \gev$, and \lumioff\ recorded at
$\sqrt{s} = 10.54 \gev$.
With an expected cross section for tau pairs at the luminosity-weighted 
$\sqrt{s}$ of $\sigma_{\tau\tau} = 0.919\pm0.003$ nb \cite{tautau},
this data sample contains about 690 million tau decays.

The \babar\ detector is described in detail in Ref.~\cite{detector}.
Charged-particle (track) momenta are measured with a 5-layer
double-sided silicon vertex tracker and a 40-layer helium-isobutane 
drift chamber inside a 1.5-T superconducting solenoidal magnet.
The transverse momentum resolution is parameterized as
$\sigma_{\pt}/\pt = (0.13\cdot \pt/[\gevc] + 0.45)\%$.
An electromagnetic calorimeter consisting of 6580 CsI(Tl) 
crystals is used to identify electrons and photons,
a ring-imaging Cherenkov detector is used to identify
charged hadrons, 
and the instrumented magnetic flux return (IFR),
embedded with limited streamer tubes and resistive plate chambers,
is used to identify muons.

A Monte Carlo (MC) simulation of lepton-flavor violating tau decays
is used to optimize the parameter space for the search.
Simulated tau-pair events including higher-order radiative
corrections are generated using \kk2f \cite{kk}
with one tau decaying to three leptons with a 3-body 
phase space distribution, while the other tau decays 
according to measured rates \cite{PDG} simulated with \tauola \cite{tauola}.
Final state radiative effects are simulated for all decays 
using \photos \cite{photos}.
The detector response is simulated with \mbox{\tt GEANT4}~\cite{geant},
and the simulated events are then reconstructed in the same 
manner as data.

The signature of the decay \taulll{} is a set of three charged 
particles, each identified as either an electron or muon,
with an invariant mass and energy equal to that of the parent 
tau lepton.
Candidate signal events in this analysis are required
to have a ``1-3 topology,'' where one tau decay yields three
charged particles, while the second tau
decay yields one charged particle.
Events with four well-reconstructed tracks 
and zero net charge are selected,
and the tracks are required to point toward a common region consistent with 
\tautau production and decay. 
The polar angle of all four tracks in the laboratory
frame is required to be within the calorimeter acceptance range.
Pairs of oppositely-charged tracks are ignored 
if their invariant mass, assuming electron mass hypotheses, 
is less than 30\mevcc, as these tracks are
likely to be from photon
conversions in the traversed material.
The event is divided into hemispheres 
in the \epem\ center-of-mass (c.m.) frame 
using the plane perpendicular to the thrust axis,
as calculated from the observed tracks and neutral energy deposits.
The signal hemisphere must contain exactly three tracks 
while the other hemisphere must contain exactly one.

Each of the charged particles found in the signal 
hemisphere must be identified as either an electron
or muon candidate. 
Electrons are identified using the ratio of
calorimeter energy to track momentum $(E/p)$, the ionization 
loss in the tracking system $(\dedx)$, and the shape of the shower
in the calorimeter.
Muon identification makes use of a neural net, 
inputs to which include the number of hits in the IFR,
the number of interaction lengths traversed,
and the energy deposition in the calorimeter.
Muons with momentum less than $500\mevc$ 
do not penetrate far enough into the IFR to be identified.
For the lepton momentum spectrum predicted by the signal MC,
the electron and muon identification requirements are found to have an
average efficiency per lepton of 91\% and 65\%, respectively.
The probability for a pion to be misidentified as an electron in 
3-prong tau decays is 2.7\%, while the probability to be misidentified
as a muon is 2.9\%.

The particle identification (PID) requirements
are not sufficient to suppress certain backgrounds, 
particularly those from light quark pair production 
and higher-order radiative 
Bhabha and \mumu events that can have four leptons 
in the final state.
To reduce these backgrounds, additional selection criteria are
applied to the six different decay modes. 
For all decay modes, the momentum of the 1-prong track is required 
to be less than 4.8 GeV/c in the c.m. frame.
Additionally, the track in the 1-prong hemisphere is assigned 
the most-likely mass hypothesis, 
and the mass of the 1-prong hemisphere 
is calculated from the four-momentum of that track
and the missing momentum in the event.
This mass is required to be in the range $0.3-3.0$ \gevcc 
for all channels except \eee\ and \eemr{}, 
for which the mass is required to be in the range $0.5-2.5$ \gevcc{}.
For the \eee\ and \eemr\ decay modes, 
radiative Bhabha events are further suppressed by rejecting
events with pairs of oppositely-charged electron tracks 
in the 3-prong hemisphere 
with invariant mass less than $250$ \mevcc{}.  
For the \eee\ and \emmr\ decay modes, the charged particle
in the 1-prong hemisphere is required to deposit energy in the
calorimeter, and must not be identified as an
electron, while for the \eemr\ and \mmm\ decay modes
this track must not be identified as a muon.
For the \eee\ and \emmr\ decay modes,
the net transverse momentum of the four tracks
must be greater than $400$\mevc{}, while for the
\eemr\ mode it must be greater than $200$\mevc{}.
Events in all six decay modes are required to have 
no track in the 3-prong hemisphere that is 
consistent with being a kaon.

To reduce backgrounds further,
candidate signal events are required to have
an invariant mass and total energy in the 3-prong
hemisphere consistent with a parent tau lepton.
These quantities are calculated from the observed track momenta 
assuming lepton masses that correspond to the specific decay mode.
The energy difference is defined as 
$\Delta E \equiv E^{\star}_{\mathrm{rec}} - E^{\star}_{\mathrm{beam}}$,
where $E^{\star}_{\mathrm{rec}}$ is the total energy of the tracks
observed in the 3-prong hemisphere and $E^{\star}_{\mathrm{beam}}$
is the beam energy, with both quantities measured in the c.m. frame.
The mass difference is defined as
$\Delta M \equiv M_{\mathrm{rec}} - m_{\tau}$ where $M_{\mathrm{rec}}$ 
is the reconstructed invariant mass of the three tracks
and $m_{\tau}=1.777\gevcc$ is the tau mass \cite{PDG}.

\begin{figure}
 \resizebox{\columnwidth}{!}{%
\includegraphics{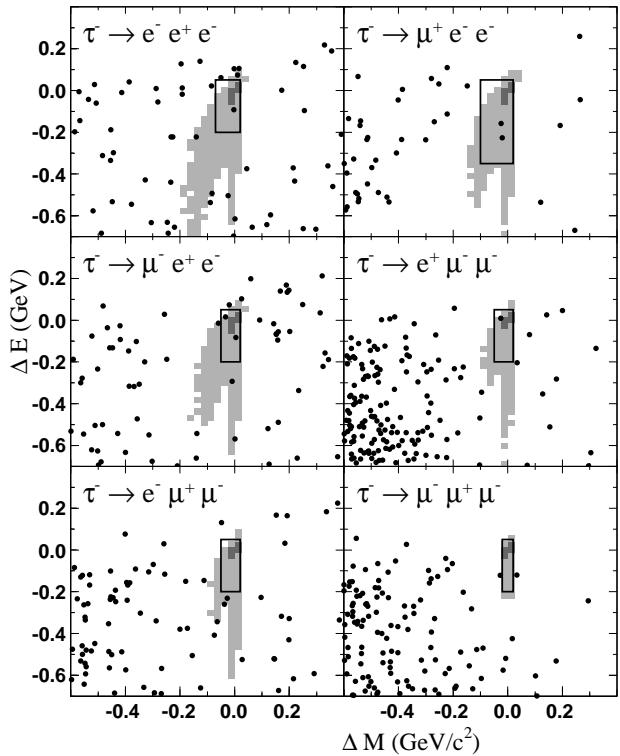}}
\caption{Observed data shown as dots in the \dEdM\ plane and 
the boundaries of the signal region for each decay mode.
The dark and light shading indicates contours containing
50\% and 90\% of the selected MC signal events, respectively.
}
\label{fig1}
\end{figure}

The signal distributions in the \dMdE\ plane (see Fig.~\ref{fig1})
are broadened by detector resolution and radiative effects.
In all decay modes, 
the radiation of photons from the incoming \epem\ particles
and from the outgoing tau particles 
leads to a tail at low values of $\Delta E$.
Radiation from the final-state leptons, 
which is more likely for electrons than muons, 
produces a tail at low values of $\Delta M$ as well.
Rectangular signal regions are defined separately for each 
decay mode.
The signal region boundaries are chosen
to provide the smallest expected upper 
limits on the branching fractions in the background-only hypothesis.
These expected upper limits are estimated using only MC
simulations and data control samples, not candidate signal events.
For all six decay modes, the upper right corner of
the signal region in the \dMdE\ plane is fixed at $(20, 50)$,
while the lower left corner is at 
$(-50, -200)$ for the \eemr\ and \emmr\ decay modes, 
$(-70, -200)$ for \eee, 
$(-100, -350)$ for \eemw, 
$(-50, -200)$ for \emmw, 
and $(-20, -200)$ for \mmm.
All values are given in units of $(\mevcc, \mev)$.
Fig.~\ref{fig1} shows the observed data in the \dMdE\ plane, 
along with the signal region boundaries 
and the expected signal distributions.
To avoid bias, a blinded analysis procedure was followed
with the number of data events in the signal region
remaining unknown until the selection criteria 
were finalized and all cross checks were performed.

There are three main classes of background remaining after
the selection criteria are applied: low multiplicity \qqbar events
(mainly continuum light-quark production); QED events (Bhabha 
and \mumu); and SM \tautau events.
These three background classes have distinctive distributions
in the \dEdM\ plane.
The \qqbar events tend to populate the plane uniformly,
while QED backgrounds are restricted to a narrow band
at positive values of $\Delta E$, and \tautau backgrounds
are restricted to negative values of both $\Delta E$ and $\Delta M$.
A negligible two-photon background remains.

The expected background rates for each decay mode are determined by
fitting a set of probability density functions (PDFs) to the
observed data in the grand sideband (GS) region of the \dEdM\ plane.
The GS region, shown in Fig.~\ref{fig1}, 
lies between $-600$ and $400$\mevcc in $\Delta M$ and
$-700$ and $400$\mev in $\Delta E$,
excluding the signal region.
For the \qqbar background, a PDF is
constructed from the product of two PDFs $P_{M^\prime}$ and %
$P_{E^\prime}$, where %
$P_{M^\prime}(\Delta M^\prime)$ is a bifurcated Gaussian and %
$P_{E^\prime}(\Delta E^\prime) = 
(1-x/\sqrt{1+x^2})(1+a x+b x^2+c x^3)$ with $x=(\Delta E^\prime-d)/e$.
The $(\Delta M^\prime, \Delta E^\prime)$ axes have
been slightly rotated from $(\Delta M, \Delta E)$ to take into
account the observed correlation between $\Delta E$ and $\Delta M$
for the distribution.
The resulting PDF has a total of eight fit 
parameters, including the rotation angle, 
all of which are determined by fits to MC \qqbar
background samples for each decay mode.
For the \tautau\ background PDF, the function $P_{M^{\prime\prime}}(\Delta M^{\prime\prime})$
is the sum of two Gaussians with common mean, while the functional form 
of $P_{E^{\prime\prime}}(\Delta E^{\prime\prime})$ is the same as that for the \qqbar\ PDF.
To properly model the wedge-shaped distribution due to the kinematic
limit in tau decays, a coordinate transformation of the form
$\Delta M^{\prime\prime}=\textrm{cos}\beta_1\Delta M+
\textrm{sin}\beta_1\Delta E$ and 
$\Delta E^{\prime\prime}=\textrm{cos}\beta_2\Delta E-
\textrm{sin}\beta_2\Delta M$ is performed. 
In total there are 11 free parameters describing this PDF, and all are
determined by fits to the MC \tautau\ samples.

For the three decay channels in which there is a significant QED background,
an analytic PDF is constructed from
the product of a Crystal Ball function \cite{CBF} in $\Delta E'$ 
and a third-order polynomial in $\Delta M'$, where again the
$(\Delta M', \Delta E')$ axes have been rotated slightly
from \dEdM\ to fit the observed distribution.
The six parameters of this PDF, including the rotation angle,
are obtained by fitting data control samples
that are enhanced in Bhabha or \mumu events.

With the shapes of the three background PDFs determined, 
an unbinned maximum likelihood fit to the data in the GS region
is used to find the expected background rate 
in the signal region, shown in Table~\ref{tab:results}.
The PDF shape determinations and background fits are performed
separately for each of the six decay modes.

The efficiency of the selection for signal events is estimated with a 
MC simulation of lepton-flavor violating tau decays.
About 40\% of the MC signal events pass the 1-3 topology requirement.
The total efficiency for signal events to be found
in the signal region is shown in Table~\ref{tab:results} for
each decay mode and ranges from 5.5\% to 12.4\%.
This efficiency includes the 85\% branching fraction for 1-prong 
tau decays.

\begin{table}
\begin{center}
\caption{Efficiency estimates, number of expected background events (\Nbgd),
expected branching fraction upper limits at 90\% CL (UL$_{90}^{\rm exp}$),
number of observed events (\Nobs), and observed branching fraction upper limits 
at 90\% CL (UL$_{90}^{\rm obs}$) for each decay mode. All upper limits are in
units of $10^{-8}$.
}
\begin{tabular}{lccccc}
\hline\hline
Mode & Eff. [\%] & \Nbgd  & UL$_{90}^{\rm exp}$ & \Nobs & UL$_{90}^{\rm obs}$ \\
\hline
\eee  &$ 8.9  \pm 0.2 $&$ 1.33 \pm 0.25 $&$ 4.9  $&$ 1 $&$ 4.3 $\\ 
\eemr &$ 8.3  \pm 0.6 $&$ 0.89 \pm 0.27 $&$ 5.0  $&$ 2 $&$ 8.0 $\\
\eemw &$ 12.4 \pm 0.8 $&$ 0.30 \pm 0.55 $&$ 2.7  $&$ 2 $&$ 5.8 $\\
\emmw &$ 8.8  \pm 0.8 $&$ 0.54 \pm 0.21 $&$ 4.6  $&$ 1 $&$ 5.6 $\\
\emmr &$ 6.2  \pm 0.5 $&$ 0.81 \pm 0.31 $&$ 6.6  $&$ 0 $&$ 3.7 $\\
\mmm  &$ 5.5  \pm 0.7 $&$ 0.33 \pm 0.19 $&$ 6.7  $&$ 0 $&$ 5.3 $\\
\hline
\hline
\end{tabular}
\label{tab:results}
\end{center}
\end{table}

The PID efficiencies and misidentification probabilities
have been measured with control samples both for data and for
MC events, as a function of particle
momentum, polar angle, and azimuthal angle in the laboratory
frame. 
The systematic uncertainties related to PID have
been estimated from the statistical uncertainties of the
efficiency measurements and from their discrepancies between data and
Monte Carlo, and range from 2.3\% for \eee\ to 12.5\% for \mmm{} \cite{uncertain}.
The modeling of the tracking efficiency contributes an additional 1\%
uncertainty.
All other sources of uncertainty in the signal efficiency are found to be 
small, including the statistical limitation of the MC signal samples,
modeling in the generator of radiative effects, 
track momentum resolution, trigger performance, observables used in 
the selection criteria, and knowledge of the tau 1-prong branching 
fractions.
The signal efficiency has been estimated using a 3-body phase space model
and no additional uncertainty is assigned for possible model dependence.
Despite the effect of slow muons on the PID efficiency, 
the total selection efficiency is found to be uniform within 
20\% across 95\% of the phase space for the three leptons.

Since the background levels are extracted directly from the data,
systematic uncertainties on the background estimation are directly
related to the background parameterization and the fit technique used.
The finite data available in the GS region to determine the
background rates contributes a significant uncertainty in all decay channels.
Uncertainties related to the background PDFs are estimated 
by varying the background shape parameters
within their errors and repeating the fits, 
and from changing the functional form of the PDFs. 
The total uncertainties on the background estimates 
are shown in Table~\ref{tab:results}.
Cross checks of the background estimation are performed by 
considering the number of events expected and observed in 
sideband regions immediately neighboring the signal region for 
each decay mode.

The numbers of events observed (\Nobs) and the 
background expectations (\Nbgd) are shown in Table~\ref{tab:results}, 
with no significant excess found in any decay mode.
Upper limits 
on the branching fractions are calculated according to 
$\BRul = \Nul/(2 \varepsilon \L \sigma_{\tau\tau})$, where $\Nul$
is the 90\% CL upper limit for the number 
of signal events when \Nobs\ events are observed with \Nbgd\ background 
events expected.
The values $\varepsilon$, $\L$, and $\sigma_{\tau\tau}$ are the
selection efficiency, luminosity, and \tautau cross section, respectively.
The uncertainty on the product $\L \sigma_{\tau\tau}$ is 1.0\%. 
The branching fraction upper limits are calculated including
all uncertainties using the technique of 
Cousins and Highland \cite{cousins92} following the implementation of 
Barlow \cite{barlow02}. 
The sensitivity or expected upper limit UL$_{90}^{\rm exp}$, 
defined as the mean upper limit expected in the background-only hypothesis, 
is included in Table~\ref{tab:results}.
The 90\% CL upper limits on the \taulll\ branching fractions 
are in the range $(4-8)\times10^{-8}$.
These limits represent up to an order of magnitude improvement 
over the previous experimental bounds \cite{babarlll, bellelll}.

\input pubboard/acknow_PRL.tex

\end{document}

%% file: pubboard/authors_jul2007.tex
%
\author{B.~Aubert}
\author{M.~Bona}
\author{D.~Boutigny}
\author{Y.~Karyotakis}
\author{J.~P.~Lees}
\author{V.~Poireau}
\author{X.~Prudent}
\author{V.~Tisserand}
\author{A.~Zghiche}
\affiliation{Laboratoire de Physique des Particules, IN2P3/CNRS et Universit\'e de Savoie, F-74941 Annecy-Le-Vieux, France }
\author{J.~Garra~Tico}
\author{E.~Grauges}
\affiliation{Universitat de Barcelona, Facultat de Fisica, Departament ECM, E-08028 Barcelona, Spain }
% ...
\collaboration{The \babar\ Collaboration}
\noaffiliation

%% file: pubboard/acknow_PRL.tex
We are grateful for the excellent luminosity and machine conditions
provided by our \pep2\ colleagues, 
and for the substantial dedicated effort from
the computing organizations that support \babar.
The collaborating institutions wish to thank 
SLAC for its support and kind hospitality. 
This work is supported by
DOE
and NSF (USA),
NSERC (Canada),
CEA and
CNRS-IN2P3
(France),
BMBF and DFG
(Germany),
INFN (Italy),
FOM (The Netherlands),
NFR (Norway),
MIST (Russia),
MEC (Spain), and
STFC (United Kingdom). 
Individuals have received support from the
Marie Curie EIF (European Union) and
the A.~P.~Sloan Foundation.

%% file: taulll_prl.bbl
\begin{thebibliography}{99}
\bibitem{cc}
Throughout this paper, charge conjugate decay modes also are implied.
\bibitem{uncertain}
All uncertainties quoted in the text are relative.
\end{thebibliography}
